\documentclass[conference]{IEEEtran}
\IEEEoverridecommandlockouts
\usepackage[a4paper,left=0.625in, right=0.625in, top=0.7in,bottom=1.07in]{geometry}
\usepackage[utf8]{inputenc}
\usepackage{cite}
\usepackage{amsmath,amssymb,amsfonts}
\usepackage{algorithmic}
\usepackage{graphicx}
\usepackage{textcomp}
\usepackage{url}
\usepackage[caption=false]{subfig}
\usepackage[justification=centering]{caption}
\usepackage[ruled,vlined]{algorithm2e}
\usepackage{epsfig}
\usepackage{float}
\usepackage{balance}
\usepackage{bbm}
\usepackage{textcomp}
\usepackage{enumitem}

\def\BibTeX{\text{B\kern-.05em{\sc i\kern-.025em b}\kern-.08em
    T\kern-.1667em\lower.7ex\hbox{E}\kern-.125emX}}

\newcommand{\beq}{\begin{equation}}
\newcommand{\eeq}{\end{equation}}

\floatstyle{ruled}
\newfloat{algorithm}{tbp}{loa}
\providecommand{\algorithmname}{Algorithm}
\floatname{algorithm}{\protect\algorithmname}

\def\BibTeX{{\rm B\kern-.05em{\sc i\kern-.025em b}\kern-.08em
    T\kern-.1667em\lower.7ex\hbox{E}\kern-.125emX}}
\begin{document}

\title{Effective Goal-oriented 6G Communications: the Energy-aware Edge Inferencing Case 
\thanks{This work has been partly funded by the European Commission through the H2020 project Hexa-X (Grant Agreement no. 101015956).}}

\author{Mattia Merluzzi$^1$, Miltiadis C. Filippou$^2$, Leonardo Gomes Baltar$^2$, Emilio Calvanese Strinati$^1$\\
$^1$CEA-Leti, Université Grenoble Alpes, F-38000 Grenoble, France\\
$^2$Intel Germany, Next Generation \& Standards, 85579 Neubiberg, Germany\\
email:\{mattia.merluzzi, emilio.calvanese-strinati\}@cea.fr,\{miltiadis.filippou, leo.baltar\}@intel.com }

\maketitle

\begin{abstract}
Currently, the world experiences an unprecedentedly increasing generation of application data, from sensor measurements to video streams, thanks to the extreme connectivity capability provided by 5G networks. Going beyond 5G technology, such data aim to be ingested by Artificial Intelligence (AI) functions instantiated in the network to facilitate informed decisions, essential for the operation of applications, such as automated driving and factory automation. Nonetheless, while computing platforms hosting Machine Learning (ML) models are ever powerful, their energy footprint is a key impeding factor towards realizing a wireless network as a sustainable intelligent platform. Focusing on a beyond 5G wireless network, overlaid by a Multi-access Edge Computing (MEC) infrastructure with inferencing capabilities, our paper tackles the problem of energy-aware \textit{dependable inference} by considering \textit{inference effectiveness} as value of a \textit{goal} that needs to be accomplished by paying the minimum price in energy consumption. Both MEC-assisted standalone and ensemble inference options are evaluated. It is shown that, for some system scenarios, goal effectiveness above 84\% is achieved and sustained even by relaxing communication reliability requirements by one decimal digit, while enjoying a device radio energy consumption reduction of almost 23\% at the same time. Also, ensemble inference is shown to improve system-wide energy efficiency and even achieve higher goal effectiveness, as compared to the standalone case for some system parameterizations.
\end{abstract}

\section{introduction}
Recently, research on the sixth generation of mobile communication systems (6G) has kicked off all around the world \cite{uusitalo2021, saad2020}, aiming to identify new technological enablers for wireless networking and service operation of unprecedented capability. Different from previous mobile systems generations, 6G will embed human, physical and digital worlds into the same ecosystem and infrastructure \cite{uusitalo2021}. To this end, both human and machine type communications will play a key role, paving the way to a holistic system in which different in-network intelligent agents exchange and process massive amounts of data in real-time to achieve common goals in several diverse scenarios (e.g. industrial, automotive). Therefore, among the several pillars needed to enable 6G, it is worth mentioning: i) a massive deployment of distributed computing and storage resources at the network edge; ii) a new communication perspective, no longer driven by radio link performance, but rather by the actual outcome of a communication strategy (\textit{effectiveness}), per a predefined goal. \\

The former is possible thanks to the Multi-access Edge Computing (MEC) paradigm \cite{kekki2018etsi}, which enables a powerful Information Technology (IT) environment including a plethora of new services at network edge in short proximity to the end user. 
The MEC paradigm poses several challenges in terms of a joint design of communication and computing 
aiming to strike the best trade-offs in terms of latency, energy, trustworthiness and, more in general, sustainability targets. Focused on Artificial Intelligence/Machine Learning (AI/ML) related workloads, an additional challenge is how to enhance learning/inference reliability given a fixed pool of radio and processing resources as parts of the network. Indeed, when considering \textit{connect-compute} services tailored to task offloading use cases (including, but not limited to learning/inference tasks) \cite{chen2018}, energy, end-to-end (E2E) latency, service dependability, etc. are affected by the availability of both radio and computing resources, whose scarcity may be severe in dense user deployments \cite{emara2021}. In this context, new measures of reliability going beyond the classical communication concepts, are ready to be explored, 
building on a paradigm shift from \textit{data-oriented} to \textit{goal-oriented communications} \cite{Letaief2022,Strinati_semantic2021}.

Goal-oriented communications go beyond the common paradigm of transmitting data with a targeted communication reliability, as focus is rather on the actual effect that the decoded information has on performing an action, a concept also known as \textit{effectiveness} \cite{Strinati_semantic2021}. While this is clear for human-to-human communications, the application to connected intelligence is a subject to be explored, paving the way to define new target performance indicators, beyond the ``legacy'' ones defining wireless technologies up to and including 5G. Such new metrics are expected to bring up new trade-offs involving several cross-layer parameters and performance.\\
\textbf{Related work:} Edge inference has recently gained considerable attention \cite{Letaief2022,Zhou19_EI}. 
In particular, data and model compression techniques have been shown to enable energy-efficient and low latency edge inference, while guaranteeing target performance in terms of accuracy \cite{Ran18,Gal21,MerluzziEML2021}, also in the case of cooperative inference \cite{ShiYua20,Merluzzi_desiree}. Goal-oriented communications have been proposed as a promising solution calling for the design of data encoding and transmission strategies, such that a specific task is performed \cite{Letaief2022,Shao2020}. In \cite{Lee19}, the authors propose a joint transmission and recognition scheme, showing the effect of communication reliability on the accuracy, ignoring delay and  energy consumption, as well as cooperative inference. Finally, task replication has been shown to improve reliability \cite{Shlezinger2021}, also in the presence of random edge server struggling or poor channel conditions~\cite{Likui20}.\\
\textbf{Our contributions:} Building on the aforementioned technological enablers, in this paper we investigate the edge inference task from a goal-oriented perspective. 
In particular, the goal embraces targeted inference accuracy/reliability and E2E delay (goal value), subject to energy consumption (goal cost), all affected by changing system parameters, such as wireless channels, MEC Hosts' (MEHs') availability, and other cross-layer parameters. Both standalone and ensemble (cooperative) inference settings are investigated, to also show the impact of inference task replication. 
The main contributions of this work are summarized as follows: i) we show how the goal-oriented approach reduces device energy consumption requirements at target effectiveness (with the latter involving delay and accuracy), even in the presence of degraded uplink communication reliability regimes (Bit Error Rate - BER); ii) we also show how a distributed deployment of MEHs enhances the success rate of the goal (effectiveness), and even further reduces the goal cost in some circumstances.\\
\textbf{Organization of the paper:} Section \ref{sec:system_model} presents the system scenario and network deployment specifics and Section \ref{sec:edge_inf_goal}  defines an edge inference goal, with its accompanied requirements and its incurred costs; Section \ref{sec:stand_vs_ensemble} elaborates upon standalone and ensemble-based inference. Section \ref{sec:perf_eval} presents thoroughly the performed numerical evaluations and observations. Finally, Section \ref{sec:conclusions} concludes the paper.

\section{System model and setup}\label{sec:system_model}
\begin{figure}[htb!]
    \centering
    \includegraphics[width=.85\columnwidth]{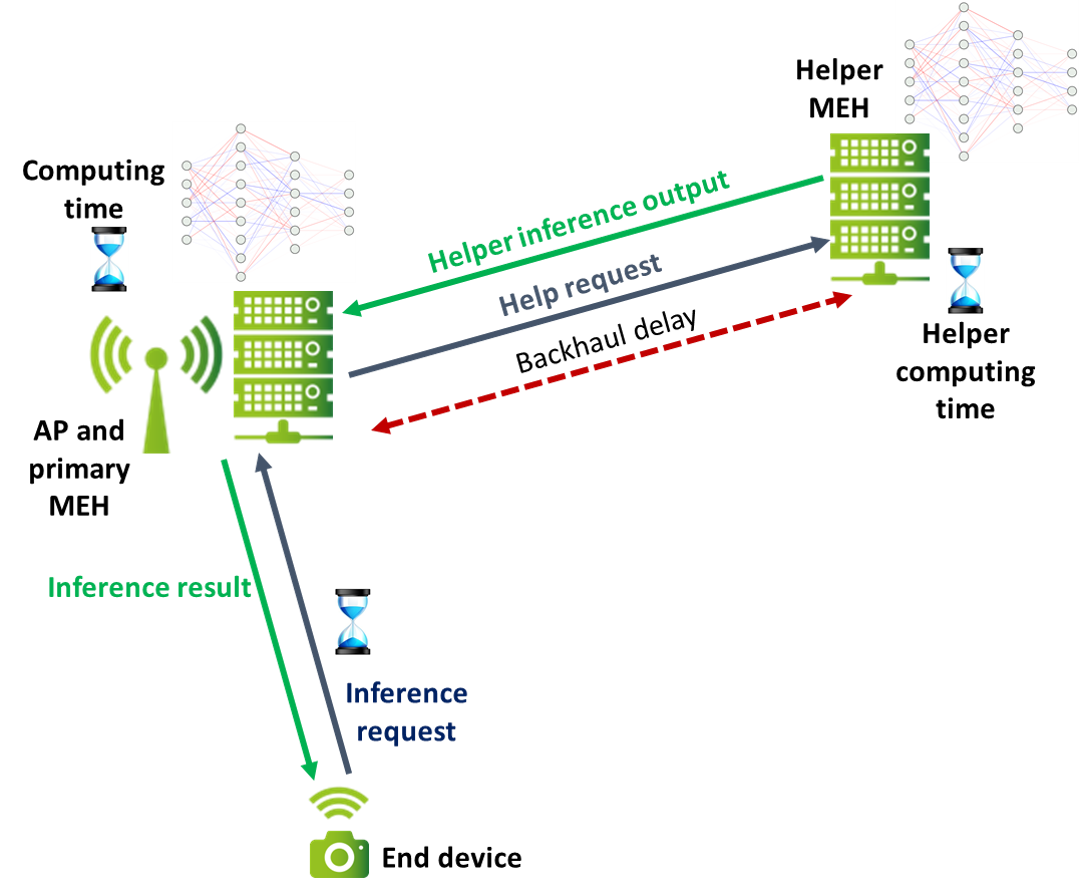}
    \caption{MEC-assisted inference scenario; use of helper MEH is optional to enhance inference performance.}
    \label{fig:network_model}
\end{figure}

The system architecture, the main system entities, and their operations are shown in Fig. \ref{fig:network_model}. In particular, in our setting, $K$ end devices (e.g., sensors or devices carrying them) of limited computing capability and power autonomy upload, through the wireless connection to an Access Point (AP), input data to nearby MEHs hosting intelligent agents, which are onboarded with inferencing task-relevant pre-trained and pre-uploaded ML models. The intelligent agents, in their turn, produce inferencing outputs (e.g., predictions or recommendations) and provide them to the calling client application via downlink radio transmission. The general assumed MEC infrastructure deployment consists of multiple MEHs deployed across a network coverage area, where, each MEH is either collocated to a radio AP, or located at aggregation points of the access network, thus, being accessible by multiple APs. For brevity, we focus on a part of a deployment consisting of two MEHs. The first one is a \textit{primary MEH}, which is collocated to the AP. 
Also, device requests can be further possibly treated by another MEH, interfaced with the primary MEH via wired (e.g., fiber) backhaul connection; this is termed after as a \textit{helper MEH}. The aim of a helper is to improve performance, e.g., in terms of goal accomplishment, i.e., inference effectiveness, thanks to offering single point of failure avoidance, as well as through inference diversity gains. 

Let us now elaborate on the goal definition, its requirements and its incurred costs for the focused system setup.

\section{Edge inference goal: Requirements \& costs}\label{sec:edge_inf_goal} 
A goal is defined as an information processing task accompanied by a set of minimum requirements that must be fulfilled as part of task accomplishment. 
In the case of edge inference, the goal is to provide inference results, within a delay budget,  overcoming a minimum tolerable level of inference accuracy. In this work, the cost of the edge inference goal is split into an energy cost for end devices to upload data via the uplink connection to an AP, and an energy cost for the MEH(s) to execute the edge inference task. To this end, we will present: i) the E2E delay model, comprising transmission signaling and computing times; ii) the goal achievement and the corresponding outages (affecting effectiveness); iii) the end devices and MEH energy consumption models. 

\subsection{End-to-end delay as a goal requirement}
Edge inference tasks fall into the category of computation offloading services, whose goal is to delegate the execution of computationally heavy workloads from end devices (e.g., vehicles or robots carrying sensors) to nearby accessible MEHs, and whose E2E delay generally entails three components for an inference task: i) uplink communication delay to transmit input inference data; ii) computation delay at the MEH to process input data; iii) downlink communication delay to transmit inference outputs (e.g., labels) back to the requesting device. In this paper, we neglect the last delay term, as in several scenarios the result represents a significantly reduced amount of bits, as compared to the input data size. 
Let us now elaborate on each delay component.
\subsubsection{Uplink transmission delay}
Let us suppose, without loss of generality, that device $k$ issues one inference request at a given time instant, and that the pattern for inference is encoded with $n_k$ bits. Thus, assuming that the transmission has to occur with target BER denoted by $\textrm{BER}_k$, the uplink transmission delay is given by:
\begin{equation}\label{delay_uplink}
    D_k^u=\frac{ n_k}{B_k^u\log_2\left(1+\frac{\varphi(\textrm{BER})h_k^up_k^u}{N_0B_k^u}\right)},
\end{equation}
where $B_k^u$ is the uplink bandwidth assigned to device $k$, $h_k^u$ is the channel power gain, $p_k^u$ is device $k$'s uplink transmit power, and $N_0$ is the noise power spectral density at the receiver; whereas, the BER margin can be approximated as $\varphi(\textrm{BER}_k)=\frac{-1.5}{\ln(5\textrm{BER}_k)}$, or $\varphi(\textrm{BER}_k)=\frac{-1.5}{\ln(0.5\textrm{BER}_k)}$ in case the spectral efficiency falls below $4$\footnote{Tighter approximations can be used under more specific assumptions on modulation and coding, which go beyond the scope of this paper.} \cite{gold1997}. 
This delay component is affected by both the device transmit power and BER requirements, a fundamental aspect in defining the difference between classical and goal-oriented communications.

\subsubsection{Remote computing delay for inferencing}
We assume that each MEH $j$ (being a primary - $p$ or a helper - $h$) is equipped with a CPU of maximum clock frequency $f_j^{\max}$. A portion of $f_j^{\max}$ is assigned to each device submitting a request for inference output. We denote by $\alpha_{k,j}\in[0,1]$ the percentage of CPU power assigned to device $k$ at MEH $j$. Furthermore, we assume that this portion is not always fully available, as MEH $j$ may dedicate part of it to higher priority traffic upon request. We assume this priority traffic to be an exogenous parameter that we cannot control, therefore, we denote by $\beta_{k,j}\in[0,1]$ the percentage of available CPU for device $k$ at a given time instant, also taking into account the concurrent arrival of higher priority processing tasks. Therefore, the actual CPU cycle frequency allocated to device $k$ by MEH $j$ is given by: 
\begin{equation}\label{f_k}
    f_{k,j}=\alpha_{k,j}\beta_{k,j} f_j^{\max}.
\end{equation}
Obviously, system performance is strongly affected by the statistics of $\beta_{k,j}$.
In the case MEH $h$ cooperates in the inference process, input inference data must be transferred from MEH $p$ to MEH $h$, and results should be then sent back to MEH $p$ to operate the necessary output aggregation. For simplicity, we assume a high capacity backhaul with fixed Round Trip Time (RTT) $D_{p,h}^{\textrm{bh}}$ between MEH $p$ and MEH $h$. 
Therefore, the total inference computation delay is
\begin{equation}\label{comp_delay_ensemble}
    D_k^c=\max\left(\frac{J_{k,p}}{f_{k,p}},\frac{J_{k,h}}{f_{k,h}}+D_{p,h}^{\textrm{bh}}\right),
\end{equation}
where $J_{k,p}$ and $J_{k,h}$ denote the numbers of CPU cycles needed to run the inference for device $k$ at MEH $p$ and $h$, respectively. Also, we neglect the time needed by MEH $p$ to aggregate results, as we assume this to be a simple operation, compared to the inference task.

Finally, the overall inference delay accounts for communication and computation delays, thus reading as $D_k^{\textrm{tot}} = D_k^u+D_k^c$. As a matter of goal accomplishment, we refer as a \textit{delay outage}, to the event $\{D_k^{\textrm{tot}}>D_k^{\max}\}$, i.e., a delay outage occurs if the total inference delay of a request exceeds a predefined threshold, required by the device during a service agreement phase. However, this represents only the first step of goal accomplishment, as from a goal perspective (\textit{correct in-time result}), an outage does not only depend on the delay of edge inference task execution, but also on the inference effectiveness, as it follows in the next section.

\subsection{Inference effectiveness and goal outage}\label{sec:goal_outage}
The \textit{inference effectiveness} can be directly represented by the accuracy of a task, or by any other metric that can quantify the success of inference operations. For example, for a classification task, it can be represented by the correct classification rate, or by the entropy computed on a posteriori probabilities at the output of a discriminative classifier. 
Instantaneously, an \textit{inference outage} occurs whenever a predefined measure does not meet a target threshold, e.g., the mean squared error in a regression task is higher than a value, or the predicted label in a classification task is incorrect. Let us refer to this measure as \textit{inference value}. 
Let us denote by $\Theta_{k,j}\in\{0,1\}$ a binary variable that equals $1$ if and only if the inference output related to a pattern generated by device $k$ and processed by MEH $j$, is accurate. 
For example, for an image classification task, $\Theta_{k,j}=1$ if an image generated by device $k$ is correctly classified. Thus, we can refer to the event $\{\Theta_{k,j}=0\}$, as \textit{inferencing outage}. 
When exploiting ensemble inference, the inference effectiveness can be improved by requesting the same inference task computation to be addressed by both MEHs, with the aim of exploiting computation and/or inference diversity. Therefore, we can write an \textit{aggregated inference value} as $\Theta_{k}=\displaystyle\Theta_{k,p}\cup\Theta_{k,h}$, where the union operator represents an aggregation of inference values. Let us note that this operation can be implemented in several ways, such as majority voting and minimum entropy in classification tasks, or any other combining strategy for different inference tasks. 
Finally, for a generic pattern generated by device $k$, a \textit{goal outage} event occurs if either a delay outage or an inference outage occurs. Therefore, we can write 
the goal accomplishment event as $\{\{\Theta_{k}=1\}\; \cap\;\{D_k^{\textrm{tot}}\leq D_k^{\max}\} \}$.
\subsection{Energy consumption as the cost of a goal}\label{sec:energy_goal}
In our scenario, we can identify two sources of energy consumption: i) the end devices, consuming energy to transmit data; ii) the MEHs, consuming energy to process data.
\subsubsection{Uplink transmission energy consumption}\label{sec:energy_UE}
We assume that the device energy consumption is only due to the uplink transmission. Therefore, recalling \eqref{delay_uplink}, the $k$-th device energy consumption for uploading a given pattern for inference is $E_k^{u}=p_k^u D_k^u$,
which, similarly to the uplink transmission delay, is affected by both transmit power and BER requirements.

\subsubsection{MEH computation energy consumption}\label{sec:energy_MEH}
The energy consumed by MEH $j$ to satisfy one inference request of device $k$ can be written as $E_{k,j} = \kappa_j f_{k,j}^2 J_{k,j}$ \cite{Yuan06}, where $\kappa_j$ is the effective switched capacitance of the processor. Thus, the total energy consumption of both involved MEHs can be written as $E_{\textrm{mec}} = \sum\nolimits_{k=1}^K\left( E_{k,p}+E_{k,h}\right)$,
where we recall that $K$ is the number of devices in the system.
\subsection{The trade-off between energy consumption \& goal outage}\label{TO_energy_goal}
The aim of this paper is to show the trade-off between average system energy consumption and goal accomplishment rate, which we term after as \textit{goal effectiveness}. The former translates into the expectation of the energy terms described in Section \ref{sec:energy_goal}, 
taken with respect to random parameters that include: i) wireless fading channels ($h_k^u,\forall k$); ii) MEH availability ($\beta_{k,i},\forall k,i$). 
The latter, recalling the goal accomplishment event presented in Section \ref{sec:goal_outage}, reads as:
\begin{equation}\label{effectiveness}
    P_k^g=\mathbb{E}\left\{\mathbf{1}\{\Theta_{k,p}=1\} \cdot\mathbf{1}\{D_k^{\textrm{tot}}\leq D_k^{\max}\} \right\},\;\; \forall k
\end{equation}
where $\mathbf{1}\{\cdot\}$ denotes the indicator function, i.e. the argument of the expectation is a Bernoulli random variable, so that its expectation is indeed the probability of the event. 
It should be noted that, in this work, we do not perform any particular optimization of the introduced parameters (e.g., power, computing resources, helper MEH selection, etc.), as the aim is an exploration of the performance aimed at providing insights on system deployment requirements, strategies, and optimization variables to be controlled during system operations. A full optimization of the system is beyond the scope of this paper and is left for future investigations. In the following section, we detail the operation of \textit{standalone} (i.e., single MEH) and \textit{ensemble inference}, from a goal effectiveness perspective.

\section{Standalone versus ensemble inference}\label{sec:stand_vs_ensemble}
We assume that a central controller at network side, which is aware of the goal requirements of a client application, collects information on channel realizations, MEHs availability, and inter-MEH backhaul delay, aiming to decide upon which MEH (MEH $p$, MEH $h$, or both) is selected to serve each incoming inference request. The central controller also provides as its output feedback (e.g., via a downlink control channel) on MEH resource availability, that can be used by each requesting device to regulate its uplink transmit power towards meeting the E2E delay requirement of the goal. \\
\textbf{Standalone inference:} Standalone inference takes place whenever MEH $p$ cannot rely on the aid of a MEH $h$. In this case, only MEH $p$ can host devices' requests, and a correct in-time inference is guaranteed only if MEH $p$ has enough computing capacity to guarantee the delay constraint (given the current connect-compute network conditions) and, at the same time, the inference is performed with the targeted inference value. 
To save device energy, we assume the device uplink transmit power to be chosen to exactly meet the delay constraint, i.e., so as to meet a transmission delay $D_{k}^u=\max\left(0,D_k^{\max}-\frac{ J_{k,p}}{f_{k,p}}\right).$
If the goal cannot be achieved with the device maximum power $p_k^{u,\max}$, we assume that the device transmits at $p_k^{u,\max}$, thus getting the inference results as soon as possible, i.e., with a best effort philosophy. \\
\textbf{Ensemble inference:} Ensemble inference takes place whenever both primary and helper MEHs cooperate in processing inferencing requests, and it can end up in the following cases, depending on the current connect-compute network conditions: i) only the primary MEH performs inference; ii) only the helper MEH performs inference; iii) both MEHs perform inference, and the result is obtained from a combination of both results (cooperative inference). 
The latter is obtained by a requesting device sending a pattern to the primary MEH, which, in its turn, sends a copy of the pattern to the helper MEH, while, in the meantime, starts processing the task; the device, in this case will await for the response of both primary and helper. This operation is feasible iff  $D_k^{u,\min} + \max\left(\frac{ J_{k,p}}{f_{k,p}},\frac{ J_{k,h}}{f_{k,h}}+D_{p,h}^{\textrm{bh}}\right)\leq D_k^{\max}$ (cf. \eqref{comp_delay_ensemble}),
where $D_k^{u,\min}$ is the minimum communication delay, obtained by plugging the maximum device transmit power $p_k^{u,\max}$ into \eqref{delay_uplink}. If this feasibility check returns a positive result, device $k$'s request is treated by means of a cooperation, choosing its transmit power so as to guarantee a transmission delay $D_{k}^u=D_k^{\max}-\max\left(\frac{ J_{k,p}}{f_{k,p}},\frac{ J_{k,h}}{f_{k,h}}+D_{p,h}^{\textrm{bh}}\right)$.
On the other hand, if cooperation is not feasible under the current conditions,  standalone inference is performed at the fastest MEH, with transmit power selected to guarantee the delay constraint, and with the best effort strategy in case of infeasibility.
\section{Performance evaluation \& discussion}\label{sec:perf_eval}
\begin{figure*}[htb!]
    \centering
    
        \subfloat[Goal effectiveness vs. device energy cons.]{
        \includegraphics[width=0.31\textwidth]{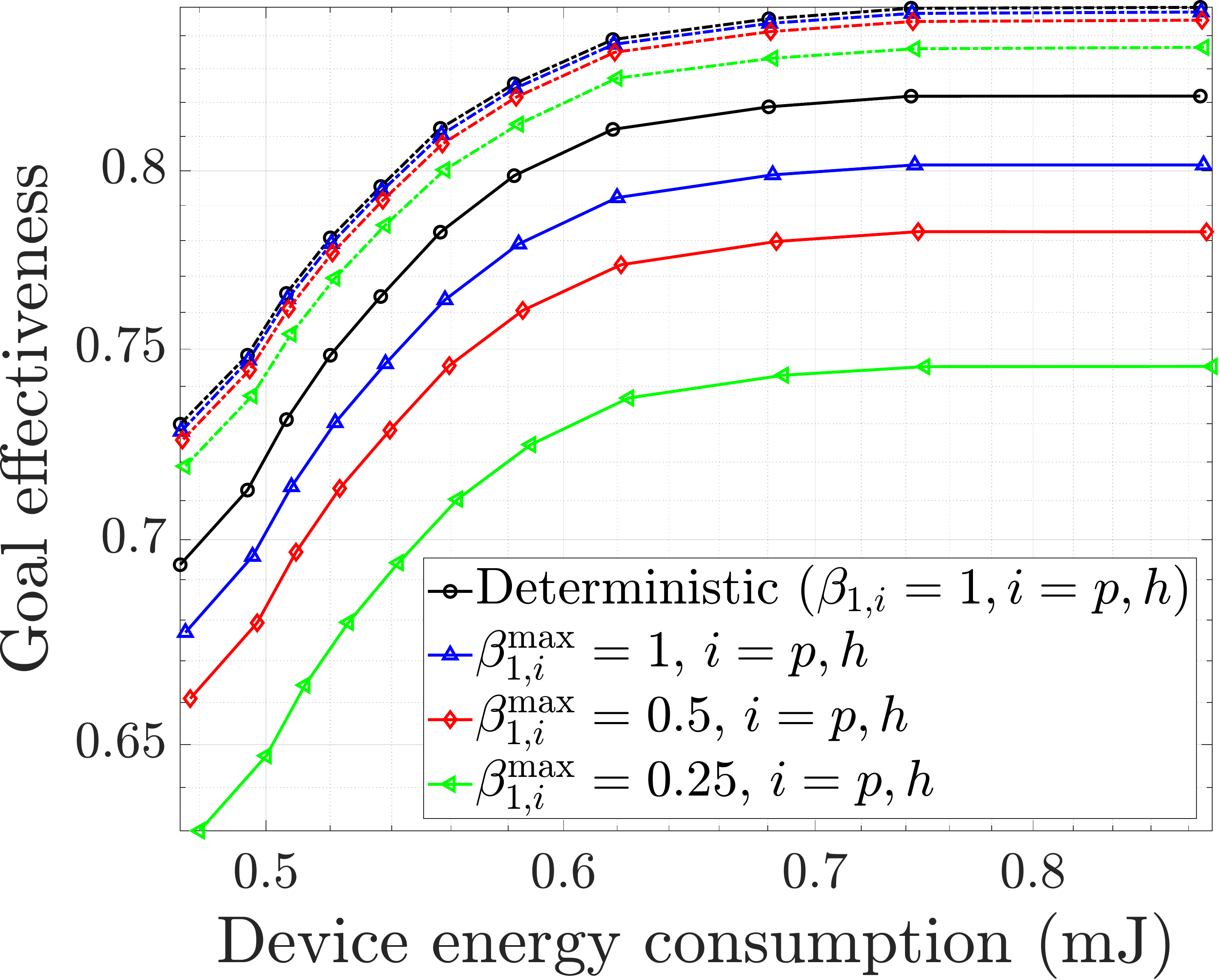}
        \label{fig:GO_vs_device_energy_ensemble}
    }
    \subfloat[Goal effectiveness vs. BER.]{
        \includegraphics[width=0.32\textwidth]{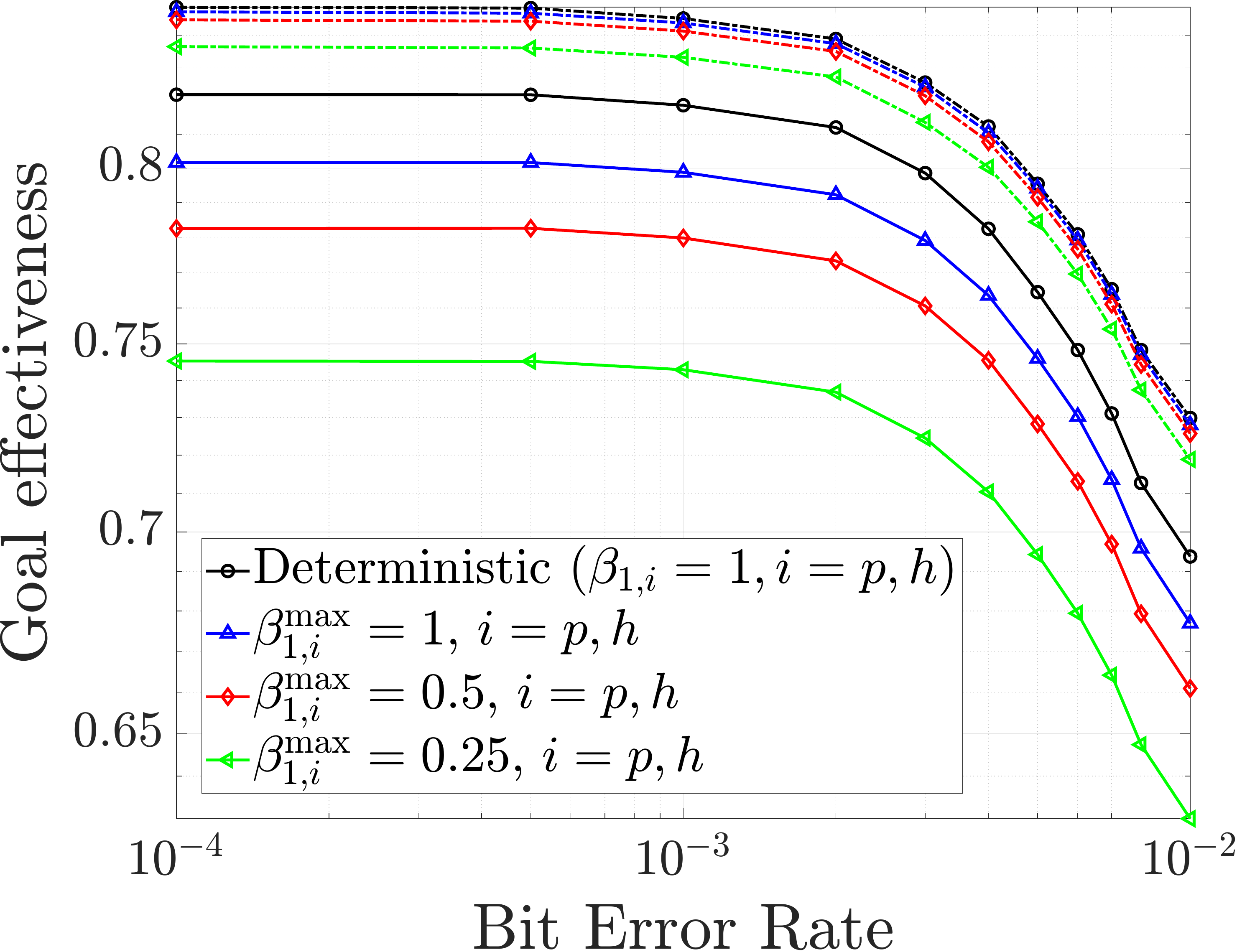}
        \label{fig:GO_vs_BER_ensemble}
    }
    \subfloat[Device energy cons. vs. BER.]{
        \includegraphics[width=.32\textwidth]{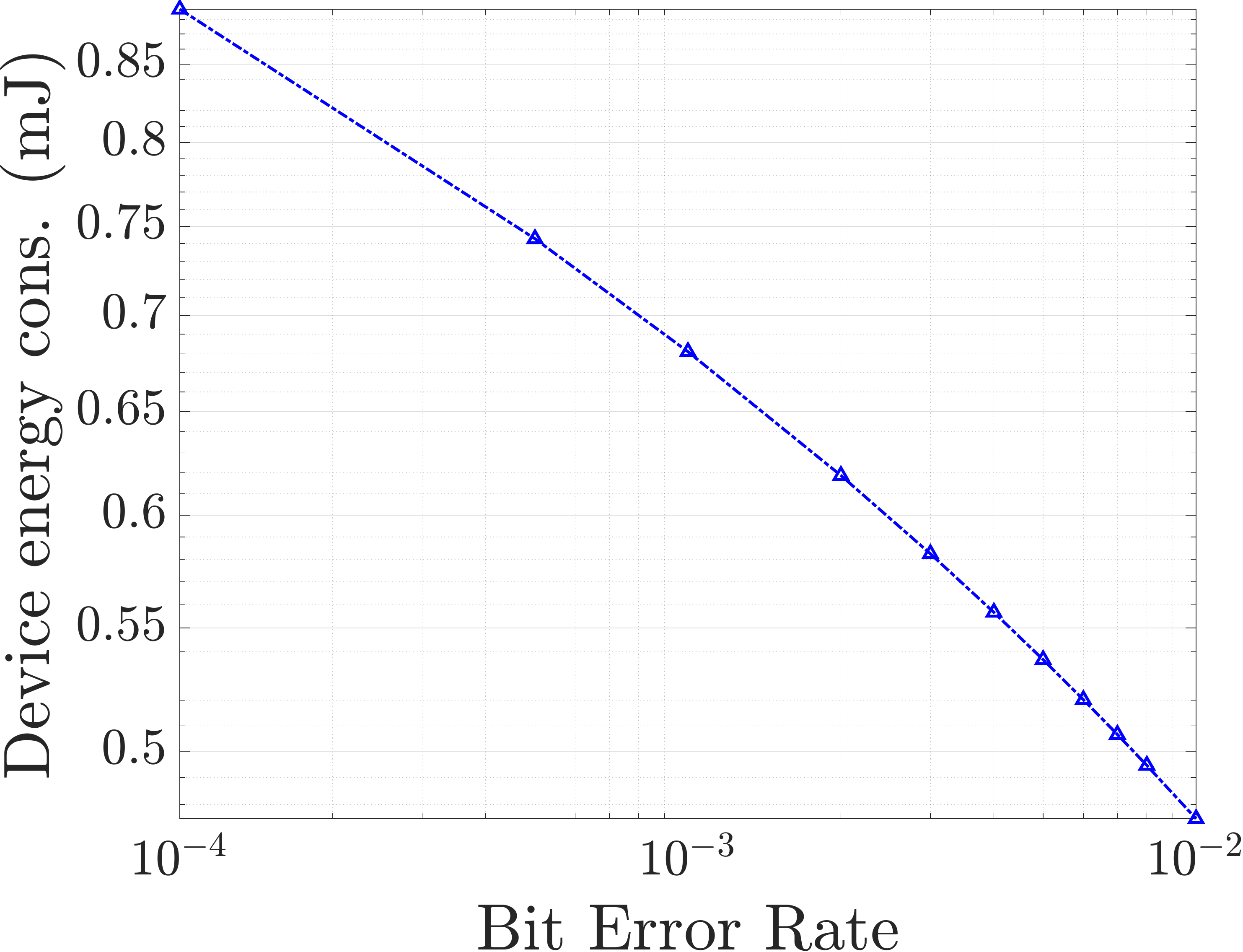}
        \label{fig:energy_vs_BER}
    }    
    
    \subfloat[MEC system energy consumption.]{
        \includegraphics[width=.318\textwidth]{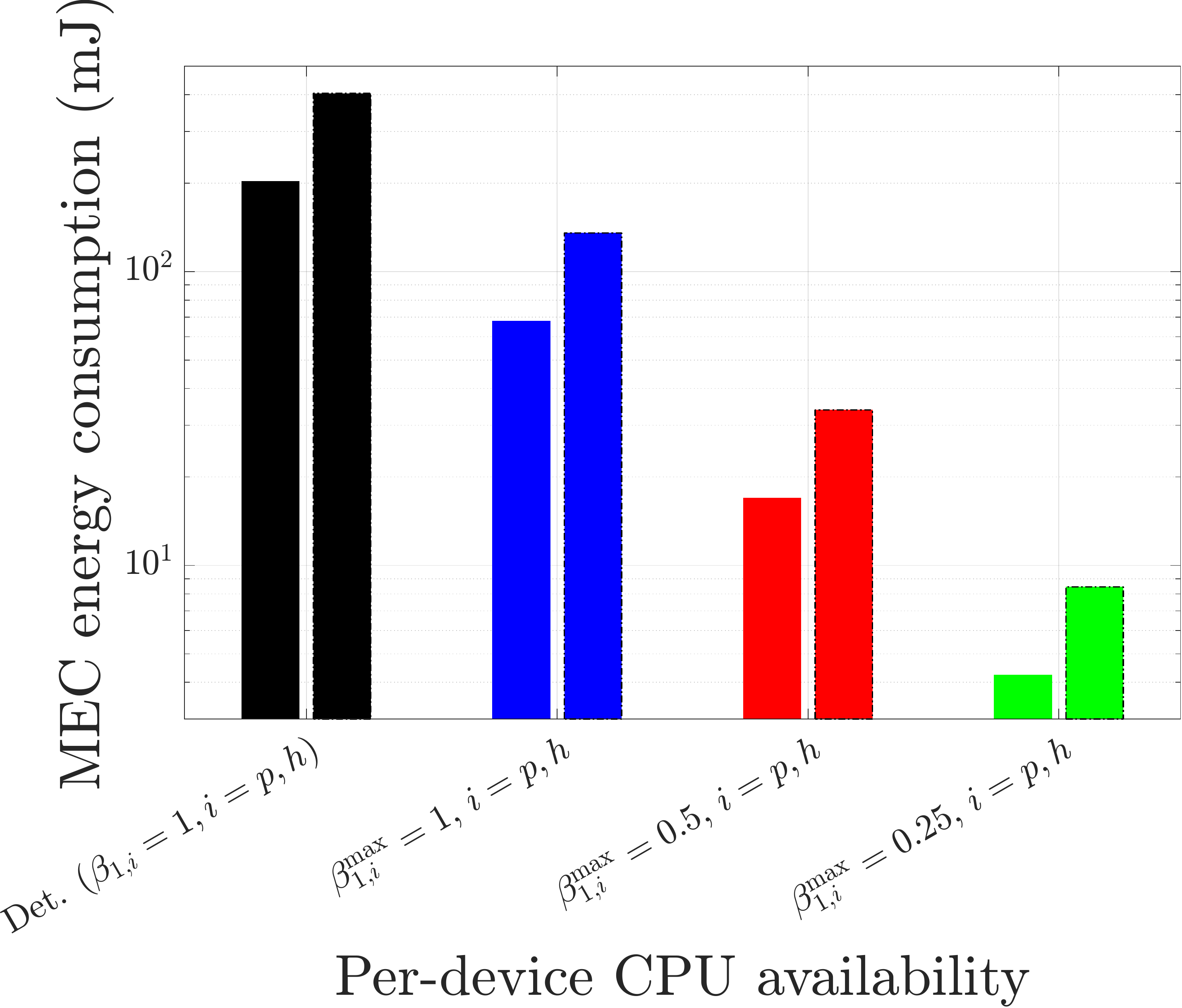}
        \label{fig:MEH_energy_ensemble}
    }
    \subfloat[Goal effectiveness for $D_{p,h}^{\textrm{bh}}=0.25D_{1}^{\max}$.]{
        \includegraphics[width=0.318\textwidth]{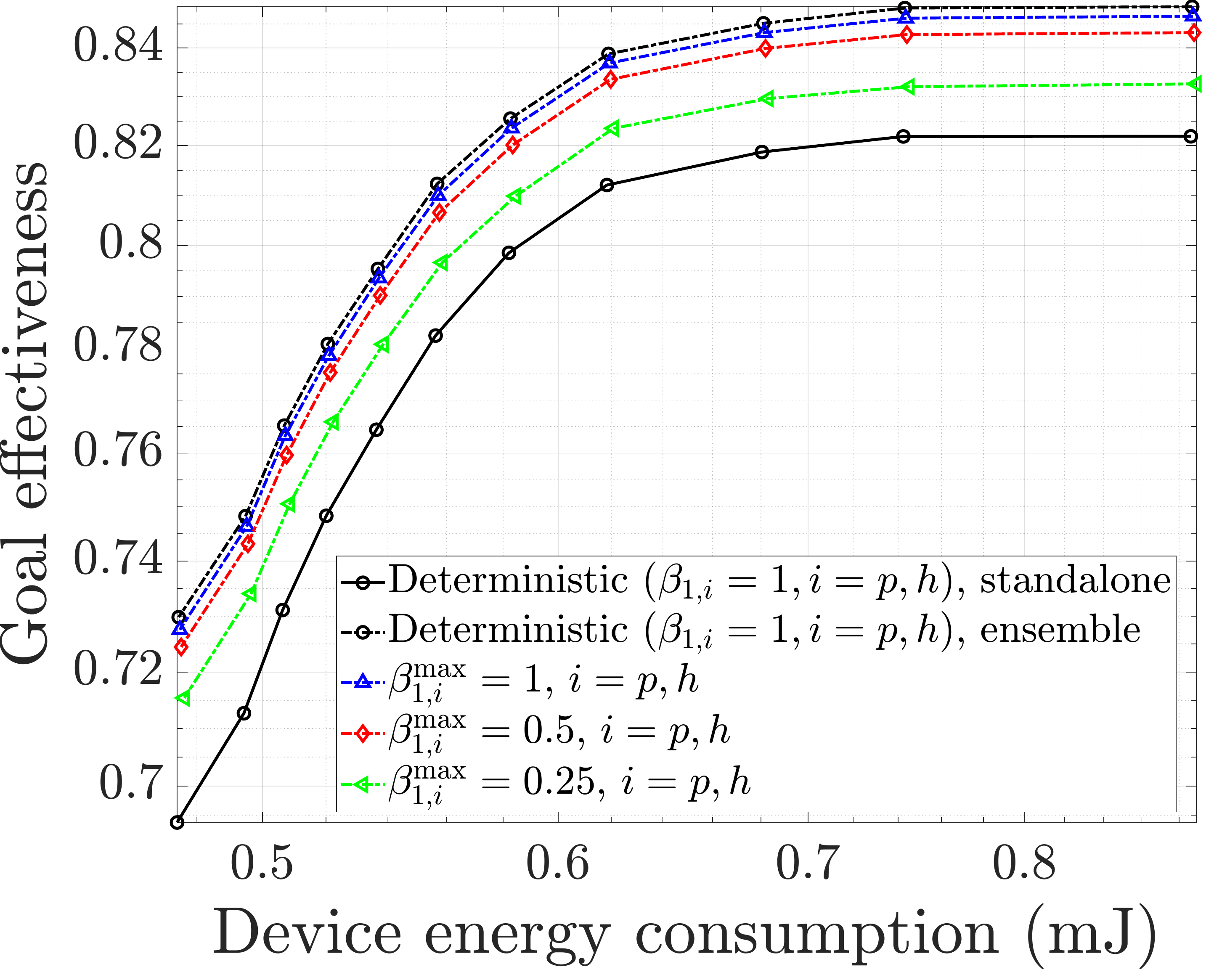}
        \label{fig:GO_bh1}
    }    
    \subfloat[Goal effectiveness for $D_{p,h}^{\textrm{bh}}=0.75D_{1}^{\max}$.]{
        \includegraphics[width=.318\textwidth]{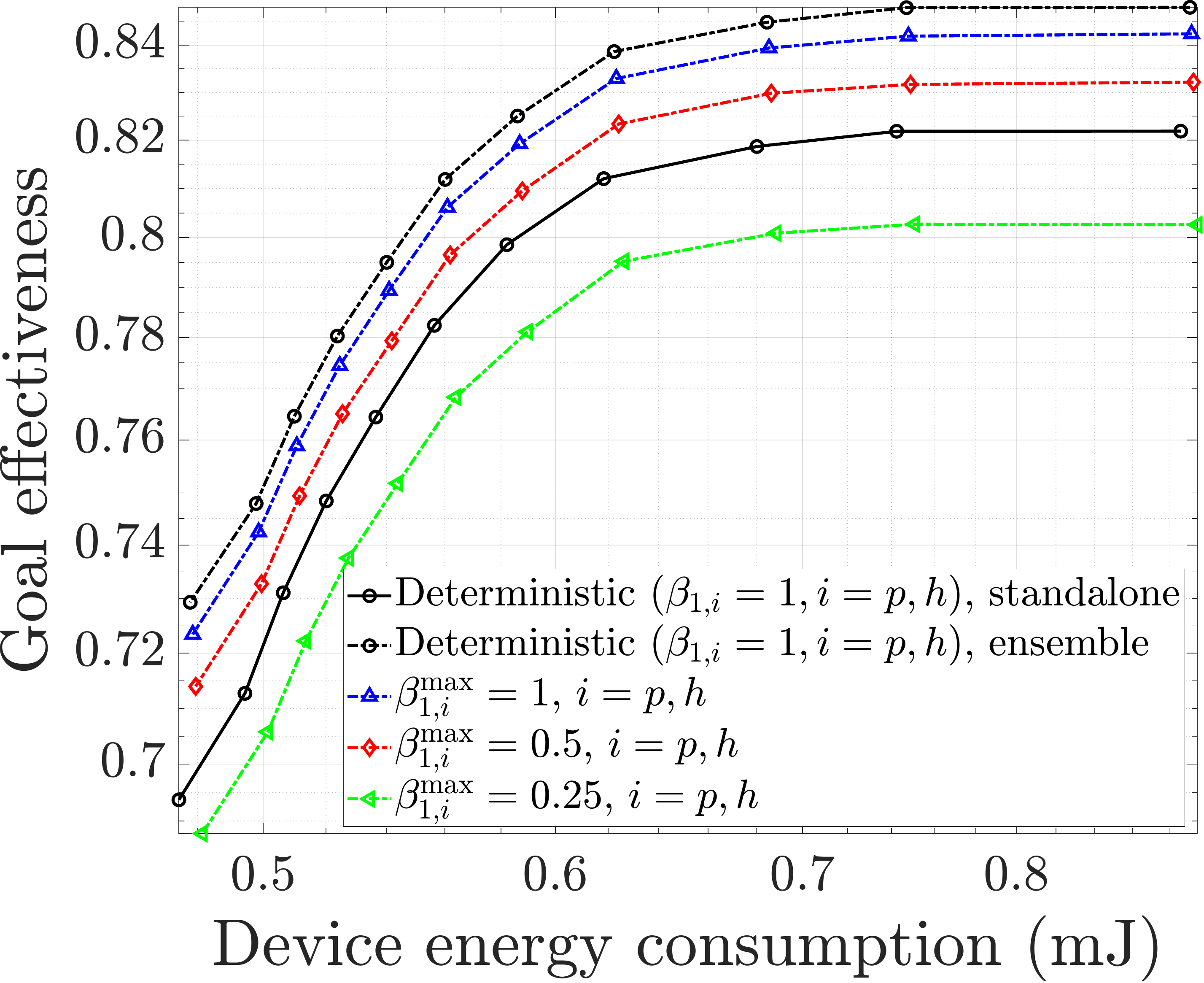}
        \label{fig:GO_bh3}
    }

    \caption{Goal achievement and system costs with different connect-compute network conditions.}
    \label{fig:perf_ensemble}
    \vspace{-.18 cm}
\end{figure*}
For the sake of performance evaluation, we focus on an edge image classification task. As indicator of inference effectiveness, we use the correct classification over a test set.\\ 
\textbf{Communication parameters:} In our performance evaluation, we consider a single device scenario, for the sake of simplicity in commenting the results. However, a multi-device scenario would not strongly affect the take home message of the simulations, which would only be influenced by the chosen task scheduling strategies at both communication and computation tiers. Device position is chosen uniformly randomly in a circle of radius $150$ meters around a single AP, serving it at $3.5$ GHz, with bandwidth of $B=10$ MHz. We generate the channels with path loss \cite{3gppchannel}, and Rayleigh fading with zero mean and unit variance. At the receiver, the noise power spectral density is $N_0=-174$ dBm/Hz, while 
the maximum device transmit power is $p_1^{u,\max} = 20$ dBm.\\
\textbf{Computation parameters:} At the MEC deployment side, we consider the primary MEH collocated with the AP, and one helper MEH, reachable within finite RTT $D_{p,h}^{\textrm{bh}}$, to be specified for each simulation. Each MEH operates at maximum CPU frequency of $f_{h}^{\max}=f_p^{\max}=4.5$ GHz. Furthermore, we assume $\beta_{1,j}, j=p,h$ (cf. \eqref{f_k}), to address device requests, taking different values to compare performance in terms of MEHs availability. Specifically, we assume $\beta_{1,p}$ and $\beta_{1,h}$ to be uniformly distributed (in three different simulations) between $0$ and $\beta_{1,j}^{\max}=[1,0.5,0.25], j=p,h$, where we also aim to capture the full (deterministic) MEH availability benchmark. The delay threshold is set to $D_1^{\max}=100$ ms.\\
\textbf{Learning parameters:} We evaluate the performance on the CIFAR-10 dataset \cite{krizhevsky2009learning}, which consists of 60000 32x32 pixel color images divided in 10 classes, with 6000 images per class. The training set is made of 50000 instances, which we split into two equal sub-data sets to train the Convolutional Neural Networks (CNNs) that are deployed at primary and helper MEHs, with the purpose of learning diversity, so as to improve inference performance in the case of cooperative inference. The test set is made of 10000 instances. We use a Deep CNN trained via usual backpropagation with an $l_2$-penalized cross-entropy loss, and an existing architecture\footnote{https://appliedmachinelearning.blog/2018/03/24/achieving-90-accuracy-in-object-recognition-task-on-cifar-10-dataset-with-keras-convolutional-neural-networks/}. Whenever exploited, the ensemble is built by summing the scores (i.e., a posteriori probabilities) of the two classifiers, and choosing the label with the highest aggregated score \cite{Shlezinger2021}.\\ 
As already mentioned in Section \ref{TO_energy_goal}, results in terms of energy consumption and goal effectiveness are obtained by averaging over realizations of wireless channels and MEH's availability. More specifically, we consider $10^5$ realizations.
\subsection{The trade-off between goal effectiveness and costs}
As a first performance evaluation, in Fig. \ref{fig:GO_vs_device_energy_ensemble}, we show the goal effectiveness (cf. \eqref{effectiveness}) as a function of the device transmit energy, obtained by varying the BER requirements during transmission. Also, to have a complete view of system performance, Figs. \ref{fig:GO_vs_BER_ensemble} and \ref{fig:energy_vs_BER} show, respectively, the goal effectiveness and the device energy consumption as functions of the same BER requirements used to obtain Fig. \ref{fig:GO_vs_device_energy_ensemble}. In these figures, the different colors encode different values of $\beta_{1,j}^{\max}$, $j=p,h$ (cf. \eqref{f_k}). At the same time, solid lines represent standalone inference (i.e., only the primary MEH participates in the inference process, without helper involvement), while dashed dotted lines represent the ensemble inference case. Let us first focus on the solid lines (i.e., standalone inference), to gain insights on goal effectiveness with respect to BER requirements (i.e., classical communication reliability) and device energy costs. Ensemble performance analysis will then come naturally as an enhancement of the standalone case, to be described in what follows below.\\
\textbf{The effect of BER on goal effectiveness:} First of all, we can  notice how the effectiveness increases, as the device transmit energy consumption increases, as expected, with different performance depending on MEH's availability levels. This is due to the fact that the BER requirements are stricter (cf. \eqref{delay_uplink}). However, we can appreciate how, above a given threshold, \textit{BER requirements do not strongly affect goal effectiveness}, meaning that a slightly increased BER leads to similar performance in terms of the effectiveness, in all network conditions (obviously with different values according to MEH availability). It is noteworthy that effectiveness is invariant to communication reliability changes in the range $\textrm{BER}=[10^{-4}$, $10^{-3}]$ (see Fig. \ref{fig:GO_vs_BER_ensemble}). This first observation (obviously dependent on the specific application) suggests that classical communication related reliability performance can be relaxed almost without any impact on goal effectiveness, and, as a by-product of such relaxation, non negligible gains in terms of device energy consumption are expected, as visible from Fig. \ref{fig:energy_vs_BER}, where the device energy consumption is plotted as a function of the BER, for $\beta_{1,i}^{\max}=1, i=p,h$. \\
\textbf{The effect of ensemble inference:} Having appreciated that, for certain system regimes, goal effectiveness may be kept almost constant by significantly relaxing classical communication related reliability, we can now analyze the beneficial effect of the helper MEH presence into the system. Indeed, from the dashed dotted lines, we can notice how goal effectiveness experiences a great improvement, if compared to the standalone inference case, on the basis of the same device energy consumption cost (i.e., the same BER requirements). Equivalently, the same effectiveness can be achieved at a lower goal cost, by \textit{further relaxing the BER requirements} (e.g., see the $80$ \% effectiveness case). \\
\indent Of course, the lower energy consumption at the device side translates into a higher energy consumption of the MEHs, as visible from Fig. \ref{fig:MEH_energy_ensemble}, where we show the energy consumption of the MEC system for a single inference request, as a function of $\beta_{1,i}, i=p,h$, in case of standalone (first bar) and ensemble inference (second bar).\\
By comparing standalone and ensemble performance for the same MEH availability, we obtain an approximately doubled energy consumption at the MEC infrastructure level for the ensemble case, as visible from Fig. \ref{fig:MEH_energy_ensemble}, due to the obvious fact that, on average, the two MEHs address requests at the same CPU speed. Nevertheless, we can notice an interesting outcome: by comparing the ensemble inference at half maximum available MEH CPU (red dashed dotted line), with the standalone inference at full maximum available CPU availability (blue solid line), better performance in terms of effectiveness against device energy consumption is guaranteed for the prior case, \textit{while also achieving a lower energy consumption at the MEC network infrastructure level}. This result is due to the quadratic law of the dynamic CPU energy consumption with respect to the CPU cycle frequency, and it shows that better performance in terms of goal achievement can be obtained without paying any price (but actually experiencing a gain), in terms of network energy consumption. In particular, this behavior suggests that, in the case of ensemble inference, it is convenient to have two available CPUs at half availability, rather than one CPU at full availability (on average), for a twofold reason: i) the total MEC energy consumption decreases; ii) the reliability improves, since single point of failure issues are counteracted. Let us also notice that, for the deterministic, the full CPU, and the half CPU availability cases,  ensemble inference outperforms the deterministic standalone method, due to the fact that the two classifiers cooperate to improve their standalone capabilities. Also, the ensemble inference at 25\% CPU availability outperforms the deterministic standalone classifier, with a greatly decreased energy consumption (around $20$ times) at the MEC infrastructure level.\\
\textbf{The effect of the backhaul delay:}  Of course, this remarkable superiority of the ensemble setting is dictated by the zero delay backhaul assumption assumed in this first simulation, which is more than ideal in practice. To this end, we now show the effect of the backahul RTT on the performance enhancement of the ensemble with respect to the standalone inference case. In Figs. \ref{fig:GO_bh1}, \ref{fig:GO_bh3}, we plot the effectiveness as a function of the device energy consumption (i.e. the same result as Fig. \ref{fig:GO_vs_device_energy_ensemble}), for two different backhaul RTT conditions, namely $D_{p,h}^{\textrm{bh}}=0.25D_{1}^{\max}$ and $D_{p,h}^{\textrm{bh}}=0.75D_{1}^{\max}$, respectively. Five MEC network conditions are considered in each figure: i) the deterministic standalone ii) the deterministic ensemble; iii) the ensemble with $\beta_{1,i}^{\max}=1, i=p,h$; iv) the ensemble with $\beta_{1,i}^{\max}=0.5, i=p,h$;  v) the ensemble with $\beta_{1,i}^{\max}=0.25, i=p,h$. From the plots, we can notice how the effect of the backhaul RTT plays a key role in determining the beneficial effects of the helper in the edge inference task, which decreases as the backhaul RTT increases. Indeed, in Fig. \ref{fig:GO_bh3}, the ensemble exhibits even degraded performance compared with the standalone, although this is due to the fact that only the deterministic standalone (i.e., its best case) is shown in the figure, which translates into a higher network energy consumption (Fig. \ref{fig:MEH_energy_ensemble}).
\section{Conclusions}\label{sec:conclusions}
In this paper, we presented the performance evaluation of an edge inference service, following a goal-oriented approach, in which system performance is not measured by classical radio reliability parameters (such as the BER), but rather 
measured by the actual outcome of a communication strategy with respect to the effectiveness of a properly defined goal. We described a MEC-enabled wireless network in which multiple MEHs can cooperate to improve edge inference performance, while trading off end device and MEC network energy as costs of the goal. 
Through numerical simulations, we have shown how the effectiveness can achieve good performance also in the presence of relaxed BER requirements, 
suggesting that high communication reliability is not needed, as far as inference performance reaches target levels. Furthermore, we have shown how ensemble inference (implemented thanks to MEHs' cooperation) can help improving goal performance, depending on specific deployment characteristics, such as the availability of MEHs and the backhaul delay interconnecting them. We believe that such a goal-oriented performance evaluation can pave the way for system deployment considerations involving communication and computing perspectives. Also, thanks to the gained insights on system performance, we hope that cross-layer optimization strategies can build on these results, to further improve trade-offs involving energy consumption and effectiveness in future wireless networks. 
\bibliographystyle{IEEEtran}
\bibliography{Ref_GO}
\end{document}